 \documentclass[final,number,3p,times,twocolumn]{elsarticle}

\usepackage{amssymb}

\usepackage{subfigure}
\usepackage{epsfig}
\usepackage[ruled,vlined]{algorithm2e}
\usepackage{verbatim}
\usepackage{graphicx}
\usepackage{multirow}

\usepackage[latin1]{inputenc} 

\biboptions{sort}

\journal{Computer Networks}

\begin{document}

\begin{frontmatter}



\title{In-packet Bloom filters: Design and networking applications}


\author[label1]{Christian Esteve Rothenberg, Carlos A. B. Macapuna, Maur\'{i}cio F. Magalh\~{a}es}
\ead{\{chesteve, macapuna, magalhaes\}@dca.fee.unicamp.br}
\author[label2]{F\'{a}bio L. Verdi}
\ead{verdi@ufscar.br}
\author[label3]{Alexander Wiesmaier}
\ead{wiesmaier@cased.de}

\address[label1]{School of Electrical and Computer Engineering 
University of Campinas, Brazil}

\address[label2]{Federal University of S\~{a}o Carlos (UFSCar), Campus Sorocaba , Brazil}

\address[label3]{Darmstadt University of Technology - CDC, 
Center for Advanced Security Research Darmstadt, Germany}

\begin{abstract}
The Bloom filter (BF) is a well-known space-efficient data structure that answers set membership queries with some probability of false positives. In an attempt to solve many of the limitations of current inter-networking architectures, some recent proposals rely on including small BFs in packet headers for routing, security, accountability or other purposes that move application states into the packets themselves. In this paper, we consider the design of such in-packet Bloom filters (iBF). Our main contributions are exploring the design space and the evaluation of a series of extensions (1) to increase the practicality and performance of iBFs, (2) to enable false-negative-free element deletion, and (3) to provide security enhancements. In addition to the theoretical estimates, extensive simulations of the multiple design parameters and implementation alternatives validate the usefulness of the extensions, providing for enhanced and novel iBF networking applications.
\end{abstract}


\end{frontmatter}


\section{Introduction}
\label{intro}
Since the seminal survey work by~\cite{Broder2005}, Bloom filters (BF)~\cite{362692} have increasingly become a fundamental data aggregation component to address performance and scalability issues of very diverse network applications, including overlay networks~\cite{1443342}, data-centric routing~\cite{citeulike:5948221}, traffic monitoring, and so on. In this work, we focus on the subset of distributed networking applications based on packet-header-size Bloom filters to share some state (information set) among network nodes. The specific state carried in the Bloom filter varies from application to application, ranging from secure credentials~\cite{1477965,Ye04statisticalen-route} to IP prefixes~\cite{1159917} and link identifiers~\cite{LIPSIN}, with the shared requirement of a fixed-size packet header data structure to efficiently verify set memberships.

Considering the constraints faced by the implementation of next generation networks (e.g., Gbps speeds, increasingly complex tasks, larger systems, high-speed memory availability, etc.), recent inter-networking proposals ~\cite{1477965, Ye04statisticalen-route, 1159917, LIPSIN, 1306762, WHITAKER02forwardingwithout} chose to include more information in the packet headers to keep pace with the increasing speed and needs of Internet-scale systems. Moving state to the packets themselves helps to alleviate system bottlenecks (e.g., IP multicast~\cite{1159917}, source routing overhead~\cite{LIPSIN}) and enables new in-network applications (e.g., security~\cite{1477965,Ye04statisticalen-route}, traceback~\cite{1306762}) or stateless protocol designs~\cite{687175}.

We refer to the BF used in this type of applications as an in-packet Bloom filter (iBF). In a way, iBFs follow a reverse approach compared to traditional standalone BF implementations: iBFs can be issued, queried, and modified by multiple network entities at packet processing time. These specific needs may benefit from additional capabilities like element removals or security enhancements. Moreover, careful BF design considerations are required to deal with the potential effects of false positives, as every packet header bit counts and the actual performance of the distributed system is a key goal.

In this paper, we address common limitations of naive iBF designs and provide a practical foundation for networking application designs requiring to solve set-membership problems on a packet basis (\S~\ref{sec:design}). Our main contribution consists of assembling and evaluating a series of practical extensions (i) to increase the system \textit{performance}, (ii) to enable false-negative-free element \textit{deletion}, and (iii) to provide \textit{security-enhanced} constructs at wire speed (\S~\ref{sec:extensions}). Via extensive simulation work, we explore the rich design space and provide a thorough evaluation of the observed trade-offs (\S~\ref{sec:evaluation}). Finally, we relate our contributions to previous work on Bloom filter designs and briefly discuss the applicability of the iBF extensions to existing applications (\S~\ref{sec:related}).

\section{Networking applications}
\label{sec:applications}
iBFs are well suited for applications where one might like to include a list of elements in every packet, but a complete list requires too much space. In these situations, a hash-based lossy representation, like a BF, can dramatically reduce space, maintaining a fixed header size, at the cost of introducing false positives when answering set-membership queries. From its original higher layer applications such as dictionaries, BFs have spanned their application domain down to hardware implementations, becoming a daily aid in network applications (e.g., routing table lookups, DPI, etc.) and future information-oriented networking proposals~\cite{SPSwitch}. As a motivation to our work and to get some practical examples of iBF usages, we first briefly survey a series of networking applications with the common theme of using small BFs carried in packets.

\subsection{Data path security}
\label{data}
The credential-based data path architecture introduced in~\cite{1477965} proposes novel network router security features. During the connection establishment phase, routers authorize a new traffic flow request and issue a set of credentials (aka capabilities) compactly represented as iBF bit positions. The flow initiator constructs the credentials by including all the router signatures into an iBF. Each router along the path checks on packet arrival for presence of its credentials, i.e., the $k$ bits resulting from hashing the packet 5-tuple IP flow identifier and the routers (secret) identity. Hence, unauthorized traffic and flow security violations can be probabilistically avoided per hop in a stateless fashion. Using 128 bits only, the iBF-based authorization token reduces the probability that attack traffic reaches its destination to a fraction of a percent for typical Internet path lengths.

\subsection{Wireless sensor networks}
\label{wireless}
A typical attack by compromised sensor nodes consists of injecting large quantities of bogus sensing reports, which, if undetected, are forwarded to the data collector(s). The statistical en-route filtering approach~\cite{Ye04statisticalen-route} proposes a detection method based on an iBF representation of the report generation (collection of keyed message authentications), that is verified probabilistically and dropped en-route in case of incorrectness. The iBF-based solution uses 64 bits only and is able to filter out 70\% of the injected bogus reports within 5 hops, and up to 90\% within 10 hops along the paths to the data sink.

\subsection{IP traceback}
\label{traceback}
The packet-marking IP traceback method proposed in~\cite{1306762} relies on iBFs to trace an attack back to its approximate source by analyzing a single packet. On packet arrival, routers insert their mark (IP mask) into the iBF, enabling a receiver to reconstruct probabilistically the packet path(s) by testing for iBF presence of neighboring router addresses.

\subsection{Loop prevention}
\label{loop}
In Icarus~\cite{WHITAKER02forwardingwithout}, a small iBF is initialized with $0$s and gets filled as forwarding elements includes adding the Bloom masks (size $m$, $k$ bits set to $1$) of the interfaces they pass to the iBF. If the {\tt OR} operation does not change the iBF, then the packet might be looping and should be dropped. If the Bloom filter changes, the packet is definitely not looping.

\subsection{IP multicast}
\label{multcast}
Revisiting the case of IP multicast, the authors of~\cite{1159917} propose inserting an iBF above the IP header to represent domain-level paths of multicast packets. After discovering the dissemination tree of a specific multicast group, the source border router searches its inter-domain routing table to find the prefixes of the group members. It then builds an 800-bit shim header by inserting the path labels ($AS_a:AS_b$) of the dissemination tree into the iBF. Forwarding routers receiving the iBF check for presence of next hop autonomous systems and forward the packet accordingly.

\subsection{Source routing \& multicast}
\label{source}
The LIPSIN~\cite{LIPSIN} forwarding fabric leverages the idea of having interface identifiers in BF-form (m-bit Link ID with only $k$ bits set to $1$). A routing iBF can then be constructed by ORing the different Link IDs representing a source route. Forwarding nodes maintain a small Link ID table whose entries are checked for presence in the routing iBF to take the forwarding decision. In addition to stateless multicast capabilities, a series of extensions are proposed to have a practical and scalable networking approach (e.g., virtual links, fast recovery, loop detection, false positive blocking, IP inter-networking, etc.). Using a 256-bit iBF in a typical WAN topology, multicast trees containing around 40 links can be constructed to reach in a stateless fashion up to 24 users while maintaining the false positive rate ($\approx  3\%$) and the associated forwarding efficiency within reasonable performance levels.

\section{Basic design} 
\label{sec:design}
The basic notation of an iBF is equivalent to the standard BF, that is an array of length $m$, $k$ independent hash functions, and number of inserted elements $n$. For the sake of generality, we refer simply to \textit{elements} as the objects identified by the iBF and queried by the network processing entities. Depending on the specific iBF networking application, elements may take different forms such as interface names, IP addresses, certificates, and so on.

On insertion, the element is hashed to the $k$ hash values and the corresponding bit positions are set to $1$. On element check, if any of the bits determined by the hash outputs is $0$, we can be sure that the element was not inserted (no false negative property). If all the $k$ bits are set to 1, we have a probabilistic argument to believe that the element was actually inserted. The case of collisions to bits set by other elements causing a non-inserted element to return ``true'' is referred to as a \textit{false positive}. In different networking applications, false positives manifest themselves with different harmful effects such as bandwidth waste, security risks, computational overhead, etc. Hence, a system design goal is to keep false positives to a minimum.

\subsection{False positive estimates}
\label{false}
It is well-known that the \textit{a priori} false positive estimate, $fpb$, is the expected false positive probability for a given set of parameters ($m$,$n$,$k$) \textit{before} element addition:
\begin{equation}
  fpb = \left[1 - \left(1 - \frac{1}{m}\right)^{k*n}\right]^k \label{Pb}
\end{equation}
The number $k$, that minimizes the false positive probability, can be obtained by setting the partial derivative of $fpb$ with respect to $k$ to 0. This is attained when $k = ln(2)*\frac{m}{n}$, and is rounded to an integer to determine the optimal number of hash functions to be used~\cite{Broder2005}.

While Eq. \ref{Pb} has been extensively used and experimentally validated as a good approximation, for small values of m the actual false positive rate can be much larger. Recently, Bose et al.~\cite{1412983} have shown that $fpb$ is actually only a lower bound, and a more accurate estimate can be obtained by formulating the problem as a balls-into-bins experiment:

\begin{equation}
  p_{k,n,m} = \frac{1}{m^{k(n+1)}} \sum_{i=1}^{m} {i^k i!} {{m}\choose{i}}  \left\{ \begin{array}{ll} 
  kn \\
  \ i \end{array} \right\}
 \label{eq:Prev}
\end{equation}

which, according to~\cite[Theorem 4]{1412983}, can be lower- and upper-bounded as follows:

\begin{equation}
  p^k < p_{k,n,m} < p^k \left( 1 + O \left( \frac{k}{n}  \sqrt{
 \frac{ln m - k ln p}{m} } \right) \right)  \label{eq:Pbound}
\end{equation}

The difference of the observed false positive rate and the theoretical estimates can be significant for small size BFs, a fact that we (and others) have empirically observed (see evaluation in \S \ref{distr_hash}). Hence, iBFs are prone to more false positives than traditional BFs for equivalent $m/n$ ratios but larger values of $m$.

Both the definition of Eq. \ref{Pb} and Eq. \ref{eq:Prev} do not involve knowing exactly how many bits are actually set to $1$. A more accurate estimate can be given once we know the fill factor $\rho$; that is the fraction of bits that are actually set to $1$ after all elements are inserted. We can define the posterior false positive estimate, $fpa$, as the expected false positive probability \textit{after} inserting the elements: 
\begin{equation}
  fpa = \rho^k \label{fpa}
\end{equation}
Finally, the observed false positive probability is the actual false positive rate ($fpr$) that is observed when a number of queries are made: 
\begin{equation}
  fpr = \frac{\mbox{Observed false positives}}{\mbox{Tested elements}}\label{fpr}
\end{equation}
Note that the $fpr$ is an experimental quantity computed via simulation or system measurements and not a theoretical estimate. Hence, the $fpr$ is the key performance indicator we want to measure in a real system, where every observed false positive will cause some form of degradation. Therefore, practitioners are less interested in the asymptotic bounds of the hash-based data structure and more concerned with the actual false positive rates, especially in the space-constrained scenario of tiny iBFs.

\subsection{Naming and operations}
\label{naming}
A nice property of hash-based data structures is that the performance is independent from the form of the inserted elements. Independently of its size or representation, every element carried in the iBF contributes with at most $k$ bits set to $1$. In order to meet the line speed requirements on iBF operations, a design recommendation is to have the elements readily in a pre-computed BF-form (m-bit vector with $k$ bits set to $1$), avoiding thereby any hashing at packet processing time. Element \textit{insertion} becomes a simple, parallelizable bitwise \texttt{OR} operation. Analogously, an iBF element \textit{check} can be performed very efficiently in parallel via fast bitwise {\tt AND} and {\tt COMP} operations.

A BF-ready element name, also commonly referred to as element \textit{footprint}, can be stored as an m-bit bit vector or, for space efficiency, it can take a \textit{sparse representation} including only the indexes of the $k$ bit positions set to $1$. In this case, each element entry requires only $ k*log_2(m)$ bits.
\section{Extensions}
\label{sec:extensions}
In this section, we describe three useful extensions to basic Bloom filter designs in order to address the following practical issues of iBFs: 
\begin{description}
\item[(i) Performance:]  \textit{Element Tags} exploit the notion of power of choices in combining hashing-based element names to select the best iBF according to some criteria, for instance, less false positives.
\item[(ii) Deletion:] \textit{Deletable Regions} introduce an additional header to code collision-free zones, enabling thereby safe (false-negative-free) element removals at an affordable packet header bit space.
\item[(iii) Security:] \textit{Secure Constructs} use packet-specific information and distributed time-based secrets to provide protection from iBF replay attacks and bit pattern analysis, preventing attackers from misusing iBFs or trying to infer the identities of the inserted elements.
\end{description}

\subsection{Element Tags}
\label{ssec:element-Tags}
The concept of {\it element Tags} (eTags) is based on extending BF-compatible element naming with a set of equivalent footprint candidates. That is, instead of each element being identified with a single footprint, every element is associated with \textit{d} alternative names, called eTags, uniformly computed by applying some system-wide mapping function (e.g., $k*d$ hash functions). That allows us to construct iBFs that can be optimized in terms of the false positive rate and/or compliance with element-specific false positive avoidance strategies. Hence, for each element, there are $d$ different eTags, where \textit{d} is a system parameter that can vary depending on the application. As we see later, a practical value of \textit{d} is in the range of multiples of 2 between 2 and 64. 

We use the notion of \textit{power of choices}~\cite{mitzenmacher-power-of-two-bf} and take advantage of the random distribution of the bits set to $1$ to select the iBF representation among the $d$ candidates that leads to a better performance given a certain optimization goal (e.g., lower fill factor, avoidance of specific false positives). This way, we follow a similar approach to the Best-of-N method applied in~\cite{jimeno}, with the main differences of (1) a distributed application scenario where the $d$ value is carried in the packet header, and (2) the best candidate selection criterion is not limited to the least amount of bits set but includes any system optimization criteria (e.g., \S~\ref{deletion} bit deletability), including those that involve counting false positives against a training set (e.g. \S~\ref{candidate} fpr-based selection).

The caveats of this extension are, first, it requires more space to store element names, and second, the value $d$ needs to be stored in the packet header as well, consuming bits that could be used for the iBF. However, knowing the $d$ value at element query time is fundamental to avoid checking multiple element representations. Upon packet arrival, the iBF and the corresponding eTag entries can be {\tt AND}ed in parallel.

\subsubsection{Generation of eTags}
\label{sssec:gen-eTags}
To achieve a near uniform distribution of 1s in the iBF, $k$ independent hash functions per eTag are required. In general, \textit{k} may be different for each eTag, allowing to adapt better to different fill factors and reducing the false positives of more sensitive elements. Using the \textit{double hashing} technique~\cite{1276233} to compute the bits set to $1$ in the $d$ eTags, only two independent hash functions are required without any increase of the asymptotic false positive probability. That is, we rely on the result of Kirsch and Mitzenmacher~\cite{1276233} on linear combination of hash functions, where two independent hash functions and can be used to simulate $i$ random hash functions of the form:
\begin{equation}
  g_i(x) = [h_1(x) + i * h_2(x)]\  mod\ m \label{eq:doublehashing}
\end{equation}
As long as $h_1(x)$ and $h_2(x)$ are system wide parameters, only sharing $i = d*k$ integers is required to derive the eTags for any set of elements. For space efficiency, another optimization for the sparse representation of the candidates consists of defining the $d$ eTags by combinations among $k + x$ iBF positions, i.e., $d = {k + x \choose k}$.


\subsubsection{Candidate selection}
\label{candidate}
Having ``equivalent'' iBF candidates enables to define a selection criteria based on some design-specific objectives. To address \textit{performance} by reducing false positives, we can select the candidate iBF that presents the best posterior false positive estimate (\textit{fpa-based selection}; Eq.~\ref{fpa}). If a reference test set is available to count for false positives, the iBF choice can be done based on the lowest observed rate (\textit{fpr-based selection}; Eq.~\ref{fpr}). Another type of selection policy can be specified to favor the candidate presenting less false positives for certain ``system-critical'' elements (\textit{element-avoidance-based selection}).

\subsubsection{False positive improvement estimate}
\label{estimate}
\begin{figure}[ht]
\centering
\subfigure[A priori false positive estimate of the iBF candidate with the lowest fill factor.]{
\includegraphics[width=0.46\textwidth]{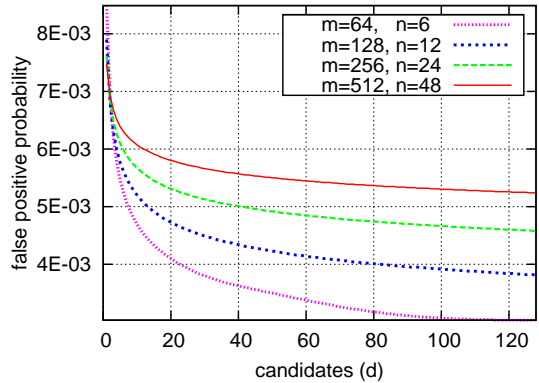}
\label{fig:fpr-d}
}
\subfigure[Potential false positive improvement when being able to choose among $d$ iBFs.]{
\includegraphics[width=0.46\textwidth]{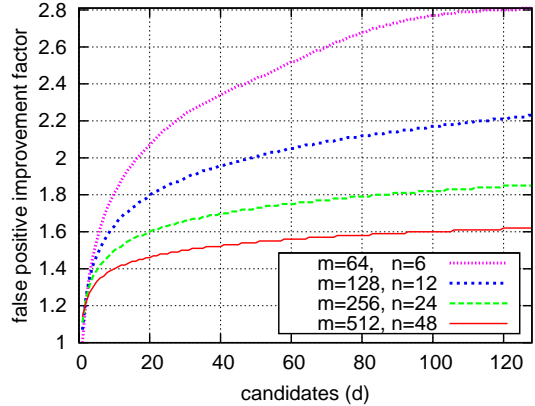}
\label{fig:fpr-d-impr}
}
\caption{False positive probability gains of the power of choices extension.
\label{fig:false-positive-d}}
\end{figure}

Following the same analysis as in~\cite{jimeno}, the potential gain in terms of false positive reduction due to selecting the iBF candidate with fewer 1s can be obtained by estimating the least number of bits set after $d$ independent random variable experiments (see~\ref{app:d-cand-fpa} for the mathematical details). Fig.~\ref{fig:false-positive-d} shows the expected gains when using the \textit{fpa-based selection} after generating $d$ candidate iBF for a given element set. With a few dozen candidates, one can expect a factor 2 improvement in the observed $fpr$ when selecting the candidate with fewer ones. Note that the four iBF configurations plotted in Fig.~\ref{fig:false-positive-d} have the same $m/n$ ratio. In line with the arguments of~\cite{1412983}, smaller bit vectors start with a slightly larger false positive estimate. However, as shown in Fig.~\ref{fig:fpr-d-impr}, the $fpr$ improvement factor of smaller iBFs due to the d-eTag extension is larger. Hence, especially for small iBFs, computing $d$ candidates can highly  improve the false positive behavior, a fact that we have validated experimentally in \S~\ref{sec:evaluation}.

\subsection{Deletable Regions}
\label{ssec:deletable-regions}
Under some circumstances, a desirable property of iBFs is to enable element deletions as the iBF packet is processed along network nodes. For instance, this is the case when some inserted elements are to be processed only once (e.g., a hop within a source route), or, when bit space to add more elements is required. Unfortunately, due to its compression nature, bit collisions hamper naive element removals unless we can tolerate introducing false negatives into the system. To overcome this limitation (with high probability), so-called counting Bloom filters (CBF)~\cite{Bonomi06animproved} were proposed to expand each bit position to a cell of $c$ bits. In a CBF, each bit vector cell acts as a counter, increased on element insertion and decreased on element removal. As long as there is no counter overflow, deletions are safe from false negatives. The main caveat is the $c$ times larger space requirements, a very high price for the tiny iBFs under consideration.

\begin{figure}[t]
\centering
\includegraphics[width=0.48\textwidth]{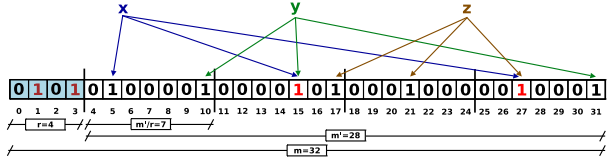} 
\caption{An example of the DlBF with $m=32$, $k=3$ and $r=4$, representing the set ${x, y, z}$. 
 The 1s in the first $r$ bits indicate that a collision happen in the corresponding region and bits therein cannot be deleted. Since each element has at least one bit in a collision-free zone, all of them are deletable.}
\label{fig:DlBF}
\end{figure}

\begin{figure}[t]
\centering
\subfigure[Deletability as function of number of regions.]{
\includegraphics[width=0.46\textwidth]{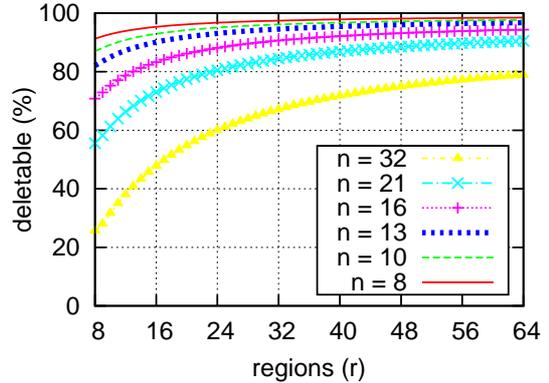}
\label{fig:del-r}
}
\subfigure[Deletability as function of inserted elements.]{
\includegraphics[width=0.46\textwidth]{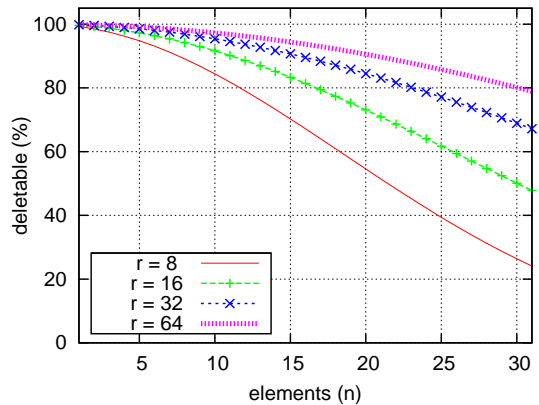}
\label{fig:del-n}
}
\caption{Deletability probability (m=256).
\label{fig:deleted}}
\end{figure}

The key idea is to keep track of where the collisions occur at element insertion time. By using the property that bits set to $1$ by just one element (collision-free bits) are safely deletable, the proposed extension consists of encoding the deletable regions as part of the iBF header. Then, an element can be effectively removed if at least one of its bits can be deleted. Again, this extension should consume a minimum of bits from the allocated iBF space. A straightforward coding scheme is to divide the iBF bit vector into $r$ regions of $m'/r$ bits each, where $m'$ is the original $m$ minus the extension header bits. As shown in Fig.~\ref{fig:DlBF}, this extension uses $r$ bits to code with 0 a collision-free region and with $1$ otherwise (non-deletable region). The probability of element deletion i.e. $n$ elements having at least one bit in a collision-free region, can be approximated to (see~\ref{app:ele-del-pro} for the mathematical details):
\begin{equation}
pdr =   1- {{n}\choose{2}}  \left( \frac{k}{r (m-r)} \right)    \label{Eq:pdr}
\end{equation}
Plotting $pdr$ against the number of regions $r$ (see Fig.~\ref{fig:del-r}) confirms the intuition that increasing $r$ results in a larger proportion of elements being deletable. As more elements are inserted into the iBF, the number of collisions increase and the deletion capabilities (bits in collision-free regions) are reduced (see Fig.~\ref{fig:del-n}). As a consequence, the target element deletion probability $pdr$ and the number of regions $r$ establish a practical limitation on the capacity of $n_{max}$ deletable iBFs.

From a performance perspective, enabling deletions comes at the cost of $r$ bits from the iBF bit space. However, removing already processed elements decreases the fill factor and consequently reduces the probability of false positives upfront. Later in \S~\ref{deletion} we explore the trade-offs between the overhead of coding the deletable regions, the impact on the $fpr$, and the implications of the candidate selection criteria.

\subsection{Secure Constructs}
\label{ssec:secure}
The hashing nature of iBFs provides some inherent security properties to obscure the identities of the inserted elements from an observer or attacker. However, we have identified a series of cases where improved security means are desirable. For instance, an attacker is able to infer, with some probability, whether two packets contain an overlapping set of elements by simple inspecting the bits set to $1$ in the iBFs. In another case, an attacker may wait and collect a large sample of iBFs to infer some common patterns of the inserted elements. In any case, if the attacker has knowledge of the complete element space (and the eTags generation scheme), he/she can certainly try a dictionary attack by testing for presence of every element and obtain a probabilistic answer to what elements are carried in a given iBF. A similar problem has been studied in~\cite{goh} to secure standalone BFs representing a summary of documents by using keyed hash functions. Our approach does not differ at the core of the solution i.e. obscuring the resulting bit patterns in the filter by using additional inputs to the hashes. However, our attention is focused to the specifics of distributed, line-speed iBF operations.

The main idea to improve the security is to bind the iBF element insertion to (1) an invariant of the packet or flow (e.g., IP 5-tuple, packet payload, etc.), and (2) system-wide time-based secret keys. Basically, the inserted elements become packet- and time-specific. Hence, an iBF gets expirable and meaningful only if used with the specific packet (or authorized packet flow), avoiding the risk of an iBF replay attack, where the iBF is placed as a header on a different packet.

\subsubsection{Binding to packet contents}
\label{packet}
We strive to provide a lightweight, bit mixing function $O = F(K, I)$ to make an element name $K$ dependent on additional in-packet information $I$. For this extension, an element name $K$ is an m-bit hash output and not the eTag representation with only $k$ bits set to $1$. The function $F$ should be fast enough to be done at packet processing time over the complete set of elements to be queried by a node processing the iBF. The output $O$ is the $k$ bit positions to be set/checked in the iBF. Using true hash functions (e.g., MD5, SHA1) as $F$ becomes unpractical if we want to avoid multiple (one per element) cycle-intense hashing per packet.

\begin{algorithm}[t]
\SetLine
\KwIn{element name $K$, packet-specific id. $I$, param. $d$}
\KwOut{$k$ bit positions to be set/checked}
\BlankLine
1.- Let $O = k \otimes I$
\BlankLine
2.- Divide $O$ into $k$ segments of $m/k$ bits:

 \hspace{0.95in}    $O_1, O_2, ..., O_k$
\BlankLine
3.- Divide each $O_j$ into a $c*log_2(m) $ bit matrix:

\hspace{\fill} $Oj_1, Oj_2, ... , Oj_{c}$ where $c = \lceil \frac{m}{k*log_2(m)}\rceil$
\BlankLine
4.- \ForEach{$O_j$ $\in [O_1, ..., O_k]$} {
   Set/check bit position $i$ in the iBF where:

    $i = O_{j1} \otimes O_{j2} \otimes ... \otimes O_{jc}$

    $i << d$
    }
\BlankLine
\caption{Secure iBF element set/check algorithm.}
\label{alg:sec-iBF}
\end{algorithm}

As an example resource-efficient implementation of $F$, we propose the lightweight Algorithm~\ref{alg:sec-iBF} to mix each element key $K$ with a fixed bit string $I$. Taking $I$ as an input, the algorithm runs in parallel on each element $K$ and returns the $k$ bit positions in the iBF to be set or checked. After an initial bitwise \texttt{XOR} operation (Step 1), the output $O$ is divided into $k$ segments of $m/k$ bits (Step 2). To build the folding matrix in Step 3, each segment is transformed into a matrix of $ c*log_2(m)$ bits.\footnote{Note that depending on the values of $m$ and $k$, some padding bits (e.g., re-used from within the segment) may be required to complete the matrix.} For instance, with $m=256$ and $k=4$, each segment $O_k$ would be a 64-bit bit vector transformed into a $8x8$ matrix. Finally, each of the $k$ output values is computed by XORing the rows of each matrix into a $log_2(m)$ bit value that returns the bit position to be set/checked (Step 4). The last d-bit shifting operation enables the power of choices capabilities.

We are faced with the classic trade-off between security and performance. An heuristic evaluation suggests that the proposed $F$ provides a good balance between performance and security. First, $F$ involves only bit shifting and {\tt XOR} operations that can be done in a few clock cycles in parallel for every $K$. Second, the $k$ bit positions depend on all the bits, within an $m/k$ bit segment, from the inputs $I$ and $K$. The security of $F$ depends on how well $I$ and $K$ are mixed. For security sensitive applications, the XOR operation in Step 1 should be replaced with a more secure transformation $P(K,I)$ i.e., using a lightweight cipher or hash function. In general, $F$ should take the application specifics into account (e.g., nature of $K$, computation of $I$ per-packet) and the target security level.

\subsubsection{Time-based keyed hashing}
\label{hashing}
A more elaborate security extension consists of using a \textit{keyed} element name construction, and change the secret key $S(t)$ regularly. We can $S(t)$ as the output of a pseudo random function $S_i=F(seed, t_i)$, where \textit{seed} is the previous seed value and $t$ a time-based input. Then, we can include the current $S$ value in the algorithm for element check/insertion e.g., $O = F(K, I, S(t))$. Thereby, we have a periodically updated, shared secret between iBF issuers and iBF processing entities, with the benefit that an iBF cannot be re-utilized after a certain period of time or after an explicit re-keying request. Moreover, by accepting $S_i$ and $S_{i-1}$ the system requires only loose synchronization similar to commercial time-coupled token generators. At the cost of initial synchronization efforts and computational overhead, this method provides an effective countermeasure to protect the system from compromised iBF attacks (cf., DDoS protection with self-routing capabilities~\cite{sec-zfilters}).

\subsection{Density factor}
\label{density}
Another iBF security measure, also proposed in~\cite{1477965}, is to limit the percentage of $1$s in the iBF to 50-75\%. A \textit{density factor} $\rho_{max}$ can safely be set to $k*n_{max}/m$, as each legitimate element contributes with at most $k$ bits. Then, the probability of an attacker guessing a bit combination causing a single false positive can be upper bounded by $\rho_{max}^k$.
\section{Practical evaluation}
\label{sec:evaluation}
In this section, we turn our attention to the practical behavior of the iBF in function of the multiple design parameters and carry out extensive simulation work to validate the usefulness of the three extensions under consideration. For these purposes, we use randomly generated bit strings as input elements and the double hashing technique using SHA1 and MD5. The section concludes exploring briefly the potential impact of different types of iBF elements (flat labels, IP addresses, dictionary entries) and the hash function implementation choice.

\subsection{Element Tags}
\label{ele_tags}
We are interested in evaluating the gains of the power of choices that underpins the element Tag extension (\S~\ref{ssec:element-Tags}), where any element set can be equivalently represented by $d$ different iBFs, different in their bit distribution but equivalent with regard to the carried element identities. We first explore the case where $k = 5$ and then the impact of using a distribution around 5 for candidate naming.\footnote{We choose $k = 5$ to have a probabilistically sufficient footprint space for the eTags ($m!/(m-k)! \approx 10^{12}$ with $m = 256$) when targeting an $m/n$ of about 8 bits per element.}
\subsubsection{Power of choices (d)}
\label{power_choice}
\begin{figure}[ht]
\centering
\subfigure[Best fill rate candidate]{
\includegraphics[width=0.46\textwidth]{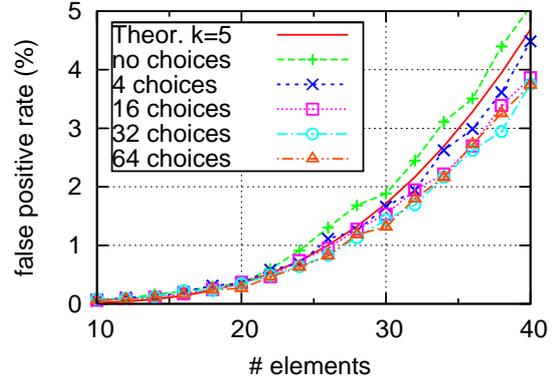}
\label{fig:ibr-fpr-D-m-256-k-5-fpa}
}
\subfigure[Best observed fpr]{
\includegraphics[width=0.46\textwidth]{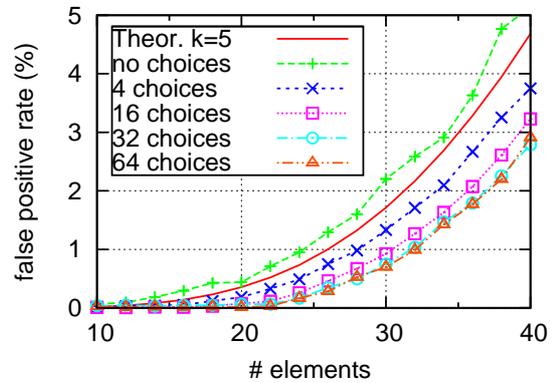}
\label{fig:ibr-fpr-D-m-256-k-5-fpr}
}
\caption{Power of choice gains (m=256, k=5).
\label{fig:ibr-fpr-D-m-256-k-5}}
\end{figure}

We run the simulations varying $d$ from 2 to 64 and updating $m$ accordingly to to reflect the overhead of including the value $d$ in the packet header. Fig.~\ref{fig:ibr-fpr-D-m-256-k-5} compares the observed $fpr$ for different values of $d$. We see that by increasing $d$ and choosing the candidate iBF just by observing its fill factor after construction (Fig.~\ref{fig:ibr-fpr-D-m-256-k-5-fpa}) leads to better performing iBFs. In the region where the iBF is more filled (30-40 elements), the observed $fpr$ drops between 30\% and 50\% when 16 or more candidate iBFs are available. Another interpretation is that for a maximal target $fpr$ we can now insert more elements. As expected, the performance gain is more significant if we consider the best performing iBF candidate after testing for false positives. Observing Fig.~\ref{fig:ibr-fpr-D-m-256-k-5-fpr}, the number of false positives is approximately halved when comparing the best iBF among 16 or more against a standard 256-bit iBF. In general, we note that the observed $fpr$ is slightly larger than the commonly assumed theoretical estimate (Eq.~\ref{Pb}), confirming thus the findings by~\cite{1412983} (Eq.~\ref{fpa}). This difference is more noticeable for small values of $m$ and becomes negligible for values larger than 1024 (see Table~\ref{tab:etag-sim-results}).

\subsubsection{Distribution of the number of hash functions (k)}
\label{distr_hash}
\begin{figure}[ht]
\centering
\subfigure[Best fill rate candidate]{
\includegraphics[width=0.46\textwidth]{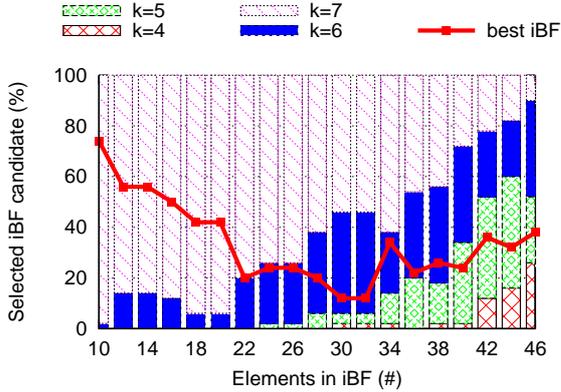}
\label{fig:ibr-fpr-kdistr-fpa}
}
\subfigure[Best observed fpr]{
\includegraphics[width=0.46\textwidth]{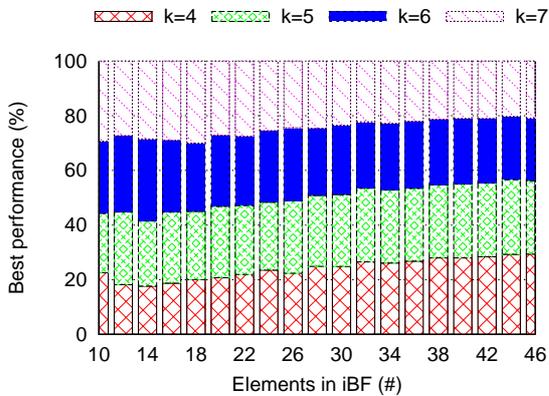}
\label{fig:ibr-fpr-kdistr-fpr}
}
\caption{Distribution of iBF candidates for different number of hash functions k. (d=16, m=256).
\label{fig:ibr-fpr-kdistr}}
\end{figure}

Now, we allow a different number of bits k per candidate. For instance, with $d=8$ the distribution of $k$ among the candidates is [4,4,5,5,6,6,7,7]. Intuitively, this naming scheme adapts better to the total number of elements in the iBF ($k$ closer to $k_{opt} = ln(2)*\frac{m}{n}$). The fpa-based selection criterion (\S~\ref{ssec:element-Tags}) is now choosing the candidate with the lowest estimate after hashing: $min{\{\rho{_0}^{k_0}, \dots  ,\rho{_d}^{k_d}\}}$. Fig.~\ref{fig:ibr-fpr-kdistr-fpa} shows the distribution of the selected 256-bit iBFs for the case of $d = 16$ and $k$ evenly distributed between 4 and 7. The line shows the percentage of times that the selected iBF actually yielded the best performance among the candidates. Disregarding the scenarios with fewer elements, the fpa-based selection criteria succeeded to choose the optimal candidate in about 30\% of the times. Fig.~\ref{fig:ibr-fpr-kdistr-fpr} shows the percentile distribution of the best performing iBF after $fpr$ testing. As expected, in more filled iBFs scenarios, setting less bits per element is beneficial. However, the differences are relatively small. As shown in Table~\ref{tab:etag-sim-results}, the observed $fpr$ in the case of $k_{const.}=5$ is practically equivalent (if not slightly better) to the case where $k$ is distributed. We can also observe what the theory in \S~\ref{sec:applications} predicts with regard to smaller iBFs and (i) their inferior $fpr$ performance for the same $m/n$ ratio, and (ii) their larger potential to benefit from the power of choices.

\begin{table}[ht]
\caption{Observed $fpr$ for iBFs with 16 eTag choices.}
\label{tab:etag-sim-results}
\begin{small}
\setlength{\tabcolsep}{5pt}
\begin{center}
\begin{tabular}{|c|c|c|c|c|c|c|c|c|}
\hline

\multicolumn{1}{|c}{\multirow{2}{*}{m} }  & \multicolumn{1}{|c}{\multirow{2}{*}{n} } & \multicolumn{2}{|c}{Std. (\%)}  & \multicolumn{2}{|c}{fpa-opt. (\%)} & \multicolumn{2}{|c|}{fpr-opt. (\%)}  \\
\cline{3-8}  &	&  Th. & $fpr$ &  $k_{cte}$ & $k_{dst}$ & $k_{cte}$ & $k_{dst}$ \\

\hline
\hline  \multirow{3}{*}{128} & 6 & 0.04 &  0.16  & 0.14  & 0.19 & 0.04  &  0.05\\
\cline{2-8}   & 12 & 0.75 & 1.12  & 0.88 & 0.86 & 0.37  & 0.32  \\
\cline{2-8}   & 18 & 3.33 & 4.39 & 2.80 & 3.10 & 2.18 & 2.37 \\

\hline
\hline  \multirow{3}{*}{256} & 12 & 0.04 &  0.09  & 0.08  & 0.08  & 0.01  &  0.03 \\
\cline{2-8}   & 24 & 0.74  & 0.95  & 0.74 & 0.71 & 0.26  & 0.30 \\
\cline{2-8}   & 36 & 3.31 & 3.63 & 2.69 & 2.75 & 2.07 & 2.15\\

\hline
\hline  \multirow{3}{*}{512} & 24 & 0.04 &  0.08  & 0.07  & 0.04  & 0.01  &  0.01\\
\cline{2-8}   & 48 & 0.74 & 0.83  & 0.64 & 0.64 & 0.22  & 0.25\\
\cline{2-8}   & 72 & 3.29 & 3.46 & 2.87 & 3.05 & 2.09 & 2.21\\

\hline
\end{tabular}
\end{center}
\end{small}
\end{table}

\subsubsection{Discussion}
\label{discussion}
Based on our experimental evaluation, having more than 32 candidates per element does not seem to bring benefits in terms of performance. If the system design choice is based on selection criteria optimized for the non-presence of specific false positives (i.e. element-avoidance \S~\ref{ssec:element-Tags}), increasing the number of choices $d$ allows complying to a larger set of false positive avoidance policies. The practical limitation appears to be solely how much space to store the candidate element representations the application designer is willing to pay.
\subsection{Deletion}
\label{deletion}
We explore two important aspects of the deletable regions extension. First, from a \textit{qualitative} point of view we examine the actual capabilities to successfully delete elements for different $m/n$ ratios, number of \textit{regions} $r$ and choices $d$. Second, we evaluate the \textit{quantitative} gains in terms of false positive reduction after element bits are deleted. Obviously, both aspects are related and intertwined with the ability to choose among candidate iBF representations to favor the deletion capabilities. Now, the application can choose the iBF candidate with the most number of bits set in collision free-zones, increasing thus the \textit{bit deletability}. Alternatively, one may want to favor the \textit{element deletability}, recalling that removing a single element bit is traduced into a practical deletion of the element.

Using our basic coding scheme (\S~\ref{ssec:deletable-regions}), we consume one bit per region to code whether collision happened and deletion is prohibited or not. Thus, the bits available for iBF construction are reduced to $m' = m - log_2{(d)} - r$.
\subsubsection{Quality}
\label{quality}
\begin{figure}[ht]
\centering
\subfigure[Deletability - elements]{
\includegraphics[width=0.46\textwidth]{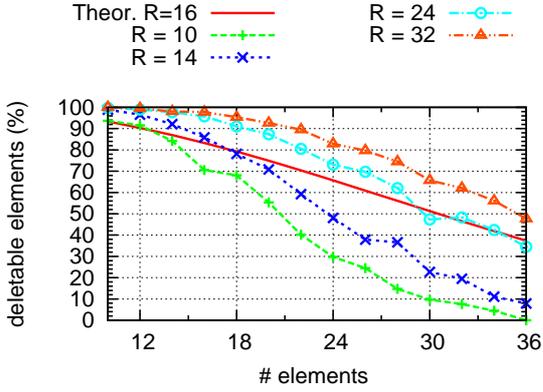}
\label{fig:Delibr-fpr-deletability-elements-D-1-m-256-k-5-R}
}
\subfigure[Deletability - bits (d=16)]{
\includegraphics[width=0.46\textwidth]{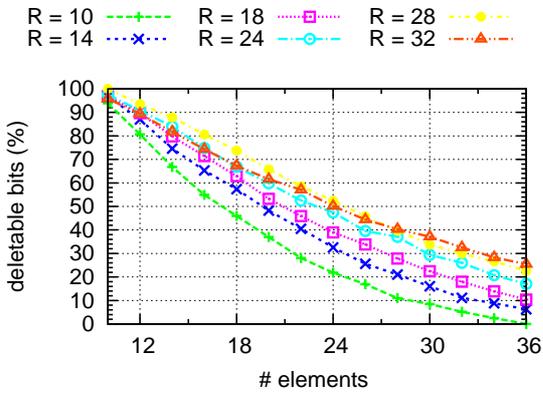}
\label{fig:Delibr-fpr-deletability-D-16-m-256-k-5-R}
}
\caption{Deletability as function of r (m=256).
\label{fig:delibr-fpr-deletability}}
\end{figure}

\begin{figure}[ht]
\centering
\subfigure[Deletability - elements]{
\includegraphics[width=0.46\textwidth]{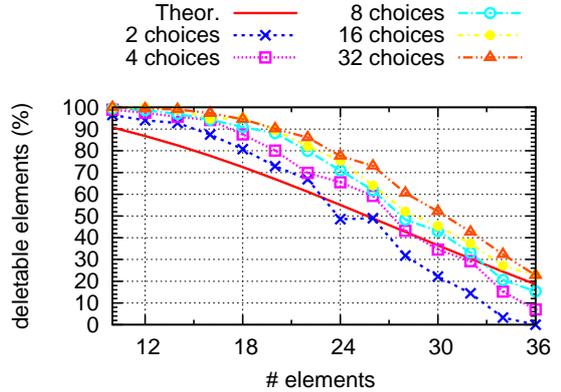}
\label{fig:delibr-fpr-deletability-elements-D-m-256-k-5-r-16-fpr}
}
\subfigure[Deletability - bits]{
\includegraphics[width=0.46\textwidth]{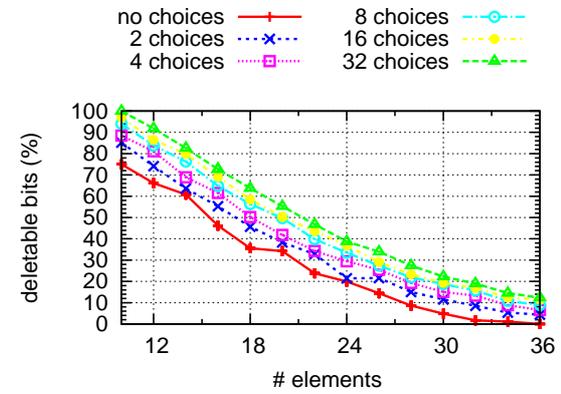}
\label{fig:delibr-fpr-deletability-D-m-256-k-5-r-16-fpr}
}
\caption{Deletability as f(d). m=256, r=16.
\label{fig:delibr-fpr-deletability-D}}
\end{figure}

We now evaluate how many of the inserted elements can be safely removed in practice. Fig.~\ref{fig:Delibr-fpr-deletability-elements-D-1-m-256-k-5-R} plots the percentage of elements that can be deleted. As expected, partitioning the iBF into more regions results in a larger fraction of elements (bits) becoming deletable. For instance, in the example of a 256-bit iBF with 32 regions (Fig.~\ref{fig:delibr-fpr-deletability}), when 24 elements are inserted, we are able to delete an average of more than 80\% of the elements by safely removing around 50\% of the bits (Fig.~\ref{fig:Delibr-fpr-deletability-D-16-m-256-k-5-R}). Playing with the candidate choices, we can enhance the bit (Fig.~\ref{fig:delibr-fpr-deletability-elements-D-m-256-k-5-r-16-fpr}) and element (Fig.~\ref{fig:delibr-fpr-deletability-D-m-256-k-5-r-16-fpr}) deletability considerably. Finally, we were able to validate experimentally the mathematical model of the element deletability probability (Eq.~\ref{Eq:pdr}).

\subsubsection{Quantity}
\label{quantity}
\begin{figure}[ht]
\centering
\subfigure[fpr - before deletion]{
\includegraphics[width=0.46\textwidth]{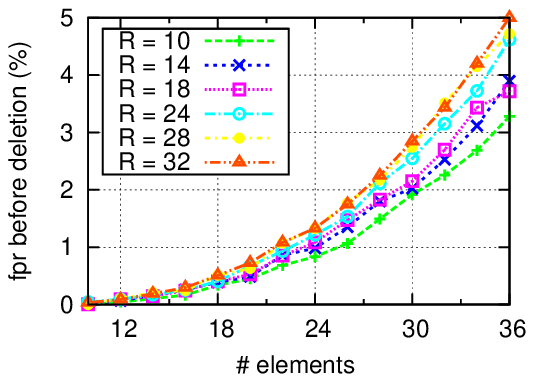}
\label{fig:Delibr-fpr-before-D-16-m-256-k-5-R}
}
\subfigure[fpr - after deletion]{
\includegraphics[width=0.46\textwidth]{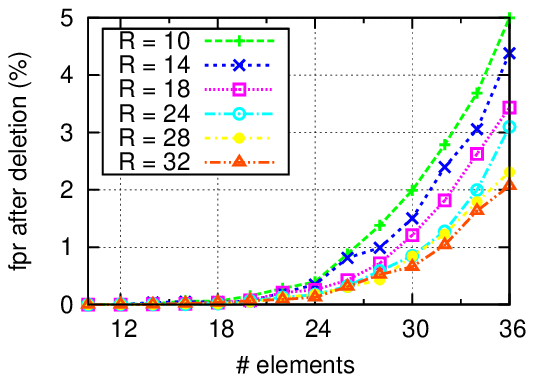}
\label{fig:Delibr-fpr-after-D-16-m-256-k-5-R}
}
\caption{False positives in function of r  (m=256, d=16).
\label{fig:delibr-fpr-regions}}
\end{figure}

Next, we explore the $fpr$ gains due to bit deletability. On one hand, we have the potential gains of removing bits from collision-free zones. On the other, the cost of (1) coding the deletable regions ($r$ bits), and (2) having more filled iBFs due to the rarefication of colliding bits. While Fig.~\ref{fig:Delibr-fpr-before-D-16-m-256-k-5-R} shows the price of having to code more regions, Fig.~\ref{fig:Delibr-fpr-after-D-16-m-256-k-5-R} illustrates the potential gains of removing evert deletable bits are removed. If we average the $fpr$ before and after elements are deleted, the iBF performance appears equivalent to the $fpr$ of a standard non-deletable m-bit iBF. In comparison, a counting BF with 2 bits per cell\footnote{Using the power of choices, we could have with very high probability a candidate that does not exceed the counter value of 3, avoiding false negatives as long as no new additions are considered.} would behave  like an iBF of size $m/2$, which would have its element capacity accordingly constrained.

\begin{figure}[ht]
\centering
\subfigure[fpr - before deletion]{
\includegraphics[width=0.46\textwidth]{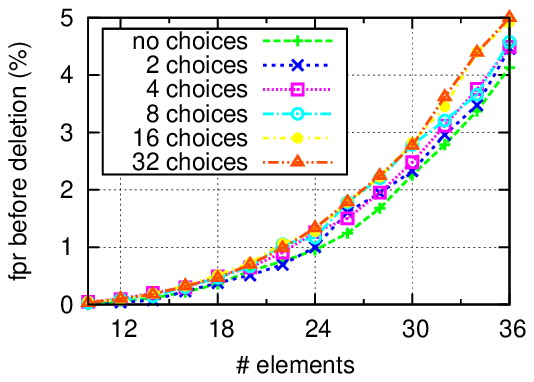}
\label{fig:Delibr-fpr-before-del-D-m-256-k-5-r-16-fpr}
}
\subfigure[fpr - after deletion]{
\includegraphics[width=0.46\textwidth]{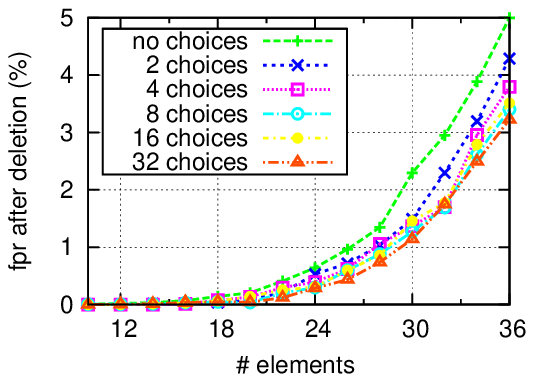}
\label{fig:Delibr-fpr-after-del-D-m-256-k-5-r-16-fpr}
}
\caption{False positives as f(d). (m=256, r=16).
\label{fig:delibr-fpr-D}}
\end{figure}

Analyzing the \textit{power of choices}, Fig.~\ref{fig:delibr-fpr-D} confirms the intuition that choosing the best deletable iBF candidate causes the colliding bits to ``thin out'' (greater $\rho$), yielding a higher $fpr$ before deletion (Fig.~\ref{fig:Delibr-fpr-before-del-D-m-256-k-5-r-16-fpr}) and a smaller $fpr$ after elements are removed (Fig.~\ref{fig:Delibr-fpr-after-del-D-m-256-k-5-r-16-fpr}).

\subsubsection{Discussion}
\label{del_Discussion}
There is a tussle between having a smaller fill factor $\rho$, with more collisions at construction time reducing the $fpa$, and the deletability extension that benefits from fewer collisions. Deletability may be a key property for some system designs, for instance, whenever an element in the iBF should be processed only once and then be removed, or when space is needed to add new elements on the fly. A more detailed evaluation requires taking the specific application dynamics (e.g., frequency of deletions/insertions) into consideration.

From a $fpr$ performance perspective, the cost of coding the deletable regions is only a slight increase in the $fpr$ due to $r$ being only a fraction of $m$. However, this practical bit space reduction seems to hinder, on average, the potential $fpr$ gains due to bit deletions upfront. Nevertheless, the deletable regions extension is a far more attractive approach to enable deletions in the space-constrained iBFs than alternative solutions based on counting BFs. An open question is whether we are able to find a more space-efficient coding scheme for the deletable regions. Finally, the power of choices again proved to be a very handy technique to deal with the probabilistic nature of hash-based data structures, enabling candidate selection for different criteria (e.g., $fpr$, element/bit deletability).
\subsection{Security}
\label{security}
Besides fast computation, the main requirements for the security extension are that (i) the random distribution of the iBF bits is conserved, and (ii) given a collection of packets~$I$ and the securely constructed iBFs, one cannot easily reveal information about the inserted elements ($K$). More generally, given a set of ($I$, $iBF$) pairs, it must be at best very expensive to retrieve information about the identities of $K$.

We first measured the randomness of the secure iBF construction outputs from Algorithm~\ref{alg:sec-iBF} by fixing a set of 20 elements and changing the per-packet 256-bit randomly generated $I$ value on each experiment run. Table~\ref{tab:eval-sec} gathers the average results of 100 experiments with 1000 runs per experiment. The observed distribution of outputs within an experiment, measured as the Hamming distance between output bit vectors (BV), was very close to the mean value of $m/2$ bits (128) with a small standard deviation.\footnote{In future work we will extend these results and the hashing techniques evaluation of Sec.~\ref{ssec:hashing_technique} with other randomness tests such as those included in the Diehard suite (http://www.stat.fsu.edu/pub/diehard).} The observed average number of bits set and their distribution were comparable to standard iBF constructs. Additionally, we analyzed whether the 20 most frequent bit positions set in secure iBFs corresponded to bits set in plain iBFs. We defined the \textit{correlation factor} as the fraction of matches and obtained a value of $0,371$, which is close to the probability of randomly guessing bits in a 256-bit iBF with $k = 4$ and $n = 20$ elements ($Pr\approx 96/256\approx 0,37$).

The results indicate that, assuming a random packet identifier $I$, first, no actual patterns can be inferred from the securely inserted elements, and second, the random bit distribution of an iBF is conserved when using the proposed algorithm. However, we recognize the limitations of Alg.~\ref{alg:sec-iBF}. For instance, if provable protection against more elaborated attacks is required, then, a more secure and computationally expensive bit mixing procedure (Step 1 in Alg.~\ref{alg:sec-iBF}) should be considered, in addition to a time-based shared secret as suggested in \S~\ref{ssec:secure}.
\begin{table}[ht]
\begin{small}
\setlength{\tabcolsep}{4pt}
\setlength{\arrayrulewidth}{1.0\arrayrulewidth}  
\begin{center}
\begin{tabular}{|l|c|c|c|}
\hline			 	&	Sec. iBF	&	Plain iBF	&	Random BV 	\\
\hline
\hline	Hamming dist. &	127.94 (8.06)	&	0	&		127.95 (8.03)	\\
\hline	\# bits set &	96.27 (3.20)	&	96.29 (-)	&		127.97 (7,97)	\\
\hline	Correlation 	&	\multicolumn{2}{|c|}{ 0.371 }	&	-		\\
\hline
\end{tabular}
\end{center}
\end{small}
\caption{Evaluation of the secure iBF algorithm (m=256, k=4, n=20). Avg. (Stdev) after 1000 runs.}
\label{tab:eval-sec}
\end{table} 
\subsection{Hashing technique}
\label{ssec:hashing_technique}
Finally, we briefly investigate the impacts of the hash function implementation choice and the nature of the input elements in small size iBFs. There are two factors that determine the ``quality'' of the bit distribution and consequently may impact the observed $fpr$: 1) the input bit string, and 2) the \textit{implementation} of the hash function.

\subsubsection{Input data sets}
\label{input_data}
Instead of considering elements as simple random bit strings, we now explore three types of elements that cover typical inputs of iBF applications: 
\begin{itemize}
\item \textbf{32-bit IP addresses:} Nearly 9M IP addresses were generated by expanding the subnet values of IP prefixes advertised in the CAIDA database.\footnote{\texttt{ftp.ripe.net/ripe/stats/delegated-ripencc-20090308
}} In addition, private IP addresses (10.0.0.0/16, 192.168.0.0/16) were also used in the experiments.
\item \textbf{256-bit random labels:} A set of 3M random labels was generated constructing each 256-bit label by picking randomly 64 hex characters and checking for uniqueness.
\item \textbf{Variable-bit dictionary words:} A set formed by 98.568 entries of the American dictionary.\footnote{\texttt{/usr/share/dict/american-english}}
\end{itemize}
\subsubsection{Hash function choice}
\label{hash_choice}
We chose 3 commonly used cryptographic hash functions (MD5, SHA1 and SHA256) and 2 general purpose hash functions (CRC32 and BOB).\footnote{Related work has investigated the properties of 25 popular hash functions, pointing to BOB as a fast alternative that yields excellent randomized outputs for network applications~\cite{1384614}. Although MD5 and SHA1 are considered broken due to the recent discovery of collisions, they are perfectly valid for our randomness purposes.}

\begin{figure}[ht]
\centering

\subfigure[Random labels - 3M \#]{ 
\includegraphics[width=0.46\textwidth]{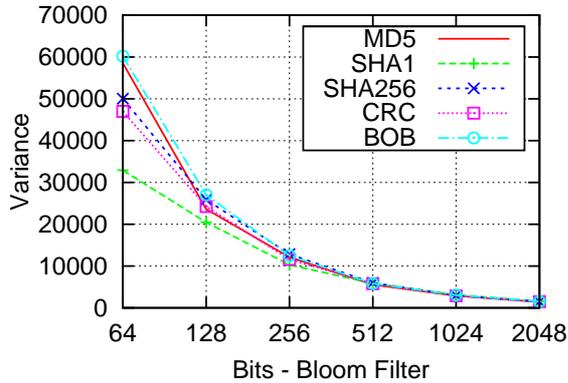}
\label{fig:av_c}
}
\subfigure[IP - 8,957,184 \#]{
\includegraphics[width=0.46\textwidth]{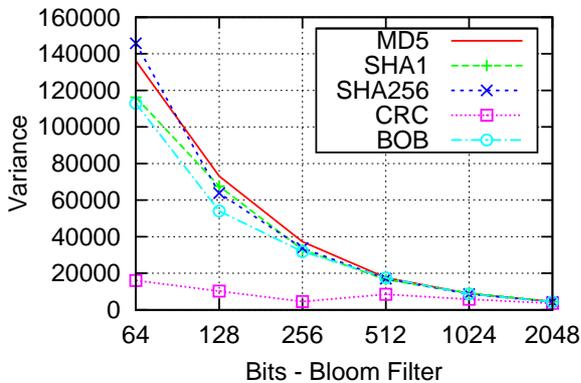}
\label{fig:av_d}
}

\caption{Normalized bit distribution variance for different hash functions.
\label{fig:aval}}
\end{figure}

The observed $fpr$ (Table~\ref{tab:doub_has}) imply that, on average, the input type does not affect the iBF performance. Figure~\ref{fig:aval} plots the observed normalized \textit{sample variance} for different bit vector sizes ($m$). For lower $m$ values the variances show a larger difference and start converging for $m > 512$. CRC presents the best output distribution when dealing with IP addresses as inputs. This may be explained by the 32-bit match of inputs and outputs. In general, the functions exhibit similar behavior, leading to the conclusion that all 5 hash functions can be used in iBF scenarios independently from the nature of the elements. This result experimentally confirms, also in the case small $m$ values, the observation by Mitzenmacher and Vadhal~\cite{1347164} that given a certain degree of randomness in the input, simple hash functions work well in practice.

\begin{table}[ht]
\begin{small}
\setlength{\tabcolsep}{3pt}
\setlength{\arrayrulewidth}{1.0\arrayrulewidth}  
\begin{center}
\begin{tabular}{|c|c|c|c|c|}
\hline $n$ & DoubleHash	& IP & Random	& Dict. 	\\
\hline

\hline	 \multirow{2}{*}{16}	&	SHA1\&MD5	&	0.340 (0.035) 	&	0.338 (0.032)	&	0.328 (0.034)	\\
\cline{2-5}	&	CRC32 segm.	&	0.345 (0.037) 	&	0.349 (0.034)	&	0.338 (0.034)	\\
\hline	\multirow{2}{*}{32}	&	SHA1\&MD5	& 2.568 (0.436) 		&	2.576 (0.449)	&	2.519 (0.385)	\\
\cline{2-5} &	CRC32 segm.	&	2.541 (0.418) 	&	2.532 (0.403)	&	2.570 (0.444)	\\
\hline
\end{tabular}
\end{center}
\end{small}
\caption{Observed $fpr$ in 256-bit iBF using double hashing with SHA1 \& MD5 and with 8-bit segments of CRC32. Avg. (StdDev); 1000 tests.}
\label{tab:doub_has}
\end{table}

\subsubsection{Hash segmentation technique}
\label{hash_seg_tec}
We note that for the purposes of iBF construction, there is a waste of hash output bits due to the $mod~m$ residual restrictions. Hence, we want to know whether we can divide the output of a hash function into $log_2(m)$ segments and use each segment as an independent hash value. We compare the bit distribution and $fpr$ performance of iBFs constructed using the double hashing technique with MD5 and SHA1 against iBFs generated with CRC32 segments as $h_i(x)$. The differences of the observed $fpr$ (Table~\ref{tab:doub_has}) are negligible, which suggests that we may indeed use this hashing technique in practice. 
 As a practical consequence, we can reduce the two independent hash function requirement of the double hashing technique to a single hash computation based on e.g., CRC32 or BOB. This result can be applied to iBF networking applications with on-line element hashing instead of pre-computed element names. Moreover, the hash segmentation technique may be useful in other multiple-hashing-based data structures (e.g., d-left hash tables) that require hashing on a packet basis.

\section{Relevance and related work}
\label{sec:related}
Although multiple variants of Bloom filter designs and applications have been proposed in the last years (e.g., \textit{Bloomier, dynamic, spectral, adaptive, retouched}, etc.), to the best of our knowledge, none of the previous work focuses on the particular requirements of distributed networking applications using small Bloom filters in packet headers.

Prior work on improved Bloom filters include the Power of Two Choices filter~\cite{mitzenmacher-power-of-two-bf} and the Partitioned Hashing~\cite{1254916}, which rely on the power of choices at hashing time to improve the performance of BFs. False positives are reduced in~\cite{1254916} by a careful choice of the group of hash functions that are well-matched to the input elements. However, this scheme is not practical in distributed, highly dynamic environments. The main idea of~\cite{mitzenmacher-power-of-two-bf} is to reduce the number 1s by choosing the ``best'' set of hash functions. Besides our in-packet-header scope, our approach differs in that we include the information of which group of hash functions was used (\textit{d} value) in the packet itself, avoiding thereby the caveat of checking multiple sets. On the other hand, we need to stick to one set of hash functions for all elements in the BF, whereas in~\cite{mitzenmacher-power-of-two-bf} the optimal group of hash functions can be chosen on an element basis. To our benefit, due to the reduced bit vector scenario, we are able to select an optimal BF after evaluating all $d$ candidates, which leads to improved performance even in very dense BF settings (small $m/n$ ratios).

Regarding the notion of choosing the best candidate filter, the Best-of-N method~\cite{jimeno} only considers a standalone application where the best BF selection is based on the least dense filter constructed with the optimal number kopt of hash functions is. Our distributed iBF applications consider candidates with different amount of bits set ($k_{distr}$) as the maximum set cardinality may be unknown and hashing at packet processing time may not be an option. Moreover, we note that distributed iBF applications may (1) be able to test for presence of elements to be queried upfront and selected the best observed $fpr$ candidate, and (2) selection criteria may be beyond reducing $fpr$, for instance benefiting the deletion of elements or avoiding specific false positives.

As far as we can say, our scheme for deleting items from a BF based on coding the regions where collisions have happened has not been proposed before. Due to its space efficiency, the false-negative-free characteristics, and the high probability of successful deletions, the Deletable Regions extension may have interesting applications beyond the scope of iBFs. The closest BF design innovation to support deletions, other than by counting BFs or d-left fingerprint hash tables, are the Variable-length Signatures (VBF) by Lu et al~\cite{Lu05bloomfilters}. While both element deletions are based on resetting at least one bit from an element signature, our scheme introduces no false negatives at the cost of providing only probabilistic guarantees for element deletion.

Security and privacy preserving extensions for BFs have been previously studied in~\cite{goh,nojima} and our main novelty resides in taking distributed systems and data packets specifics into consideration (e.g., flow-identifier, time-based loose synchronization of distributed secrets).

Recently, our work on iBFs has contributed to enable a novel packet forwarding scheme with built-in DDoS protection~\cite{sec-zfilters}. The notion of element Tags has its roots in our work on the link identifier based forwarding fabric in~\cite{LIPSIN}, rendering the system more useful (network policy compliance, loop avoidance, security) and efficient ($fpr$ control, larger multicast groups).
 Similarly, the d-eTag extensions may be applied to the case of IP multicast~\cite{1159917} to reduce false positives and compliance to inter-domain AS policies in the case of false positives. When applied to the credentials based architecture proposed in~\cite{1477965}, multiple candidates may allow iBFs to transverse larger paths before reaching the maximum density. Moreover, the security extension may provide extra protection from an en-route attacker spoofing the source IP address and re-using the flow credentials for unauthorized traffic. Finally, the hash segmentation technique appears useful to lower the burden on networking elements when computing multiple hashes on a packet basis.
\section{Conclusions}
\label{conclusions}
This paper explores an exciting front in the Bloom filter research space, namely the special category of small Bloom filters carried in packet headers. Using iBFs is a promising approach for networking application designers choosing to move application state to the packets themselves. At the expense of some false positives, fixed-size iBFs are amenable to hardware and present a way for new networking applications.

We studied the design space of iBFs in depth and evaluated new ways to enrich iBF-based networking applications without sacrificing the Bloom filter simplicity. First, the power of choices extension shows to be a very powerful and handy technique to deal with the probabilistic nature of hash-based data structures, providing finer control over false positives and enabling compliance to system policies and design optimization goals. Second, the space-efficient element deletion technique provides an important (probabilistic) capability without the overhead of existing solutions like counting Bloom filters and avoiding the limitations of false-negative-prone BF extensions. Third, security extensions were considered to couple iBFs to time and packet contents, providing a method to secure iBFs against tampering and replay attacks. Finally, we have validated the extensions in a rich simulation set-up, including useful recommendations for efficient hashing implementations. We hope that this paper motivates the design of more iBF extensions and new networking applications.

\section*{Acknowledgments}
We want to thank the reviewers of earlier versions of this paper. Specially, we appreciate the fruitful discussions with Mats Naslund, Pekka Nikander and Andras Zahemszky. The work presented in this paper is supported by Ericsson Research, CNPq and FAPESP.

\section*{References}
\bibliographystyle{elsarticle-num} 
\bibliography{ibf}







\appendix
\section{Mathematical Model for Element Deletability Probability}
\label{app:ele-del-pro}
Consider a vector of size $m$, $k$ hash functions and $r$ regions. The bit vector size to construct the actual BF is $m' = m - r$ and the number of bit cells in each region is equal to $\lceil m' / r \rceil$. 
 The probability of two hash functions setting a given bit cell:
\begin{small}
\begin{equation}
  p_{1} =   \left( \frac{1}{m'} \right)  \left( \frac{1}{m'} \right) = \left(  \frac{1}{m'^{2}} \right)   \label{Eq:pdr1}
\end{equation}
\end{small}

If we use one hash function per element (i.e. one bit) and insert $n$ elements, the probability of having at least one collision in a given cell is:
\begin{footnotesize}
\begin{eqnarray}
  p_{2} =  {{n}\choose{2}}  \left( \frac{1}{m'^2} \right) + {{n}\choose{3}}  \left( \frac{1}{m'^3} \right)
+ \cdots + {{n}\choose{n}}  \left( \frac{1}{m'^n} \right) = \sum_{i=2}^{n} {{n}\choose{i}}  \left( \frac{1}{m'^{i}} \right)   \label{Eq:pdr2}
\end{eqnarray}
\end{footnotesize}
%
Considering that each element insertion sets $k$ bits to $1$, we compute the probability to:
\begin{footnotesize}
\begin{eqnarray}
  p_{3} =  {{n}\choose{2}}  \left( \frac{b_1}{m'^2} \right) + {{n}\choose{2}}  \left( \frac{b_2}{m'^2} \right)
+ \cdots + {{n}\choose{2}}  \left( \frac{b_k}{m'^2} \right)  = \sum_{i=2}^{n} {{n}\choose{i}}  \left( \frac{k}{m'^{i}} \right)   \label{Eq:pdr3} 
\end{eqnarray}
\end{footnotesize}
%
Equation \ref{Eq:pdr3} considers a collision in a given bit cell. Now, we extend the probability to the $m' / r$ bit cells in a single region:
\begin{small}
\begin{equation}
  p_{4} = \left( 1-p_{3} \right)^{\frac{m'}{r}}   =  \left( 1 - \sum_{i=2}^{n} {{n}\choose{i}}  \left( \frac{k}{m'^{i}} \right) \right)^{\frac{m'}{r}}   \label{Eq:pdr4}
\end{equation}
\end{small}
Replacing $m'$ with $m-r$, the probability of an element being deletable, that is of having at least one bit set in a collision-free region is:
%
\begin{small}
\begin{equation}
  pdr  =  \left( 1 - \sum_{i=2}^{n} {{n}\choose{i}}  \left( \frac{k}{{m-r}^{2}} \right) \right)^{\left( \frac{m-r}{r} \right)}   \label{Eq:pdr6}
\end{equation}
\end{small}
Finally, neglecting the contributions of the terms for $i > 2 $ as they tend to zero, we get:

\begin{small}
\begin{equation}
pdr \approx  \left( 1 - {{n}\choose{2}}  \left( \frac{k}{{m-r}^{2}} \right) \right)^{\left( \frac{m-r}{r} \right)}  
\label{Eq:pdr-approx}
\end{equation}
\end{small}

\section{Mathematical model for d-candidate fpa optimization (adapted from~\cite{jimeno})}
\label{app:d-cand-fpa}

Given the iBF parameters $m$, $n$, $k$, and letting $d$ be the number of different iBF candidates for the same element set, the probability of setting $s$ bits in an iBF candidate can be formulated as an independent random variable experiment:
\begin{footnotesize}
\begin{equation}
E^{2} \left[ s \right] = m \left( 1- \left( 1- \frac{1}{m} \right)^{kn}\right) + m \left(m-1 \right) \left( 1-2 \left( 1- \frac{1}{m} \right)^{kn} + \left( 1- \frac{2}{m} \right)^{kn} \right)
\label{Eq:fpa1}
\end{equation}
\end{footnotesize}

\begin{footnotesize}
\begin{equation}
\sigma^{2} \left[ s \right] = m \left( \frac{m-1}{m} \right)^{kn} + m^2 \left( \frac{m-2}{m} \right)^{kn} + m \left( \frac{m-2}{m} \right)^{kn} + m^2 \left( \frac{m-1}{m} \right)^{2kn}
\label{Eq:fpa2}
\end{equation}
\end{footnotesize}

Defining $\mu = E[s]$ and $\sigma = \sigma[s]$, the minimum continuous probability density function is: 

\begin{small}
\begin{equation}
f_{min} \left( s \right) = \left( \frac{1}{2^{d-1}} \right) d \left( erfc \left( \frac{s-\mu}{\sigma \sqrt{2} } \right) \right)^{d-1} \left( \frac{1}{\sigma \sqrt{2 \pi} } e^{\frac{-\left(s-\mu \right)}{2 \sigma^2}  } \right)
\label{Eq:fpa3}
\end{equation}
\end{small}

Consequently, the expectation of the least number bits ($s_{min}$) set by any of $d$ candidates:

\begin{small}
\begin{equation}
E\left(S_{min} \right) = \int_{-\infty}^{\infty}{s f_{min} \left(s \right) ds }
\label{Eq:fpa4}
\end{equation}
\end{small}

Finally, the probability of a false positive once the smallest fill ratio has been estimated:

\begin{small}
\begin{equation}
pr \left[false positive \right]= \left( \frac{E\left[s_{min} \right] }{m}^k \right)
\label{Eq:pr}
\end{equation}
\end{small}

\end{document}